# Shannon Mutual Information Applied to Genetic Systems

J.S. Glasenapp: Optical Sciences Center, University of Arizona, Tucson 85721 - U.S., glasenapp@optics.arizona.edu
B.R. Frieden: Optical Science Center, University of Arizona, Tucson 85721 - U.S.
C.D. Cruz: Departamento de Biologia Geral, Universidade Federal de Viçosa, Viçosa 36570 - Brasil

**Abstract**

Shannon information has, in the past, been applied to quantify the genetic diversity of many natural populations. Here, we apply the Shannon concept to *consecutive generations* of alleles as they evolve over time. We suppose a genetic system analogous to the *discrete noisy channel* of Shannon, where the signal emitted by the input (mother population) is a number of alleles that will form the next generation (offspring). The alleles received at a given generation are conditional upon the previous generation. Knowledge of this conditional probability law allows us to track the evolution of the allele entropies and mutual information values from one generation to the next. We apply these laws to numerical computer simulations and to real data (*Stryphnodendron adstringens*). We find that, due to the genetic sampling process, in the absence of new mutations the mutual information increases between generations toward a maximum value. Lastly, after sufficient generations the system has a level of mutual information equal to the entropy (diversity) that it had at the beginning of the process (mother population). This implies no increase in genetic diversity in the absence of new mutations. Now, obviously, mutations are essential to the evolution of species. In addition, we observe that when a population shows at least a low level of genetic diversity, the highest values of mutual information between the generations occurs when the system is neither too orderly nor too disorderly. We also find that mutual information is a valid measure of allele fixation.

# 1   Introduction

In genetics, there are two complementary branches of knowledge: The neutral theory of molecular evolution (Kimura) and natural selection theory (Darwin). In general, the two theories are regarded as "mutually exclusive".  In this paper, we assume Kimura's neutral theory. In comparison, Kimura's neutral theory allows the study of evolution at the molecular level, in part because it provides a way to make strong predictions that can be tested against field data. Kimura's theory holds that most variation at the molecular level does not affect fitness. On this basis, neutral mutations are effectively 'silent' in their effects upon fitness. In addition, Kimura's theory is stochastic in nature, this allow us to describe the genetic variation by the statistics-based theory of information called "Shannon information theory".

Currently Shannon information theory [1] has been applied to several scientific fields within their own context. In cellular biology, it was used to detect complexity functional, activity and secondary structure of enzymes and to measure heterogeneities of genomes [2-4]. Tkačik *et al*., [5] have shown that maximizing the mutual information between a transcription factor and its target also maximizes the positional information of the cells in the embryo. This defines the anterior-posterior position in the fruit fly (*Drosophila*). Fath *et al*., [6] describes how various measures of information, including Shannon's, are thought to contribute to the survival and evolution of species.

In ecology [7] and population genetics [8] Shannon's entropy equation is been used as a diversity index [9-13]. Since it has the requirements that a measure of genetic diversity must have: It is minimum (in fact zero) when there is only one allele in one locus; it is a maximum for equal allele frequencies; it increases as the number of alleles increases, and it is a convex function of the allele frequencies [8]. However, the most useful measures of genetic diversity are based on the theory of random genetic drift [14] and the neutral theory of molecular evolution [15].  According to Kimura's neutral mutations model, a new mutant allele spreads across the population by chance, which means by genetic drift. The mutations are the only source of "new" genetic variation and all existing molecular polymorphism [16]. Mutations that do not result in a different amino acid being replaced in the chain are called *silent* mutations, and mutations that result in different amino acid being replaced in the chain are called *replacement* mutations [14]. Natural selection, even of very modest intensity, will have a critical effect on both the direction and speed of evolutionary process, particularly in large populations. However, in Kimura's view, only a small portion of DNA changes in the evolution is adaptive in nature [16]. Disadvantageous mutations are usually eliminated by *negative* selection; nearly all other mutations are supposed neutral or almost so [14].

Although, a study between different *Drosophila* species estimate 25% of divergence in replacement sites caused by positive selection, in fact, the confidence interval estimated is huge (45% or as low as 5%). Johnson [17], argued that even though the highest percentage challenge in some ways the neutral theory, the lower number is compatible to Kimura's neutral theory. Despite the number of substitutions high as 45% driven by positive selection, the number of substitutions is far fewer than the total number of mutations. Studying the genetic variation in the gene addresses coat colors among wild and domestic pigs from both Europe and Asia, Fang *et al*. [18] found a similar number of mutations in wild and domestic pigs. Nevertheless, the nature of those mutations were different. All mutations found among wild boars were silent. Therefore, is still seems clear that the majority of the new mutations are either neutral or deleterious, with only a relatively small fraction being advantageous.

Indeed, it is not the purpose of this article to discuss the plausibility of the neutral theory. Although positive selection appears to be more important than Kimura thought, the neutral model is still regarded as a major contributor to molecular evolution. Neutral theory enable researchers to establish new research programs, including the generation of many tests used to demonstrate selection [17].

Our goal is provide a new tool suitable for applying to genetic diversity studies of natural populations, which depend as well upon dominant or codominant DNA markers and isozymes. The neutral nature of data coming from molecular markers, allow us to combine Shannon information theory with Kimura's neutral theory in order to describe the flow of genetic information from one generation to the next. Therefore, it is possible applying information theory to population genetic studies as complementary approach to the rich evolutionary theory already existing. Here, as in Kimura's model, we assume that the factor causing fluctuations in gene frequency is simply the random sampling of gametes [15].

Lewontin [8] showed that the Shannon definition [1] of entropy (e.g. see Eq. (4) below) is a suitable measure of genetic diversity in a hierarchically divided population. Here, we propose a new approach by applying Shannon's equations to get measures of genetic diversity and mainly, the mutual information $I(x,y)$ between *consecutive generations* as they evolve over time. This is a mutual 'intergenerational' information. We suppose a genetic system, which undergoes a transition from an input state ($\alpha$) to an output state ($\beta$). Here a system of alleles, representing an input state ($\alpha$) at the generation time $t$-1 are transmitted into an output state ($\beta$) at the next generation time $t$. Note that to determine the mutual information does not require knowledge of what the source is of the fluctuations in the allele frequencies between generations. We model the transitions of alleles like the transitions of symbols over the *noisy-channel* of the communication system of Shannon. Thus, an allele received at generation $t$ randomly differs from the allele 'sent' at generation $t$ -1. We show that this is possible by using a special matrix which is composed by the conditional probabilities $P(y_j|x_i)$ of receiving alleles $y_j$ at generation $t$ given that alleles $x_i$ were sent at generation $t$ -1. The indices $i$ and $j$ range independently over all allele possibilities in one locus.

We apply these ideas to computer simulations and to real data from a medicinal plant (*Stryphnodendron adstringens*). We compute measures of genetic diversity and mutual information $I(x,y)$ flow between successive generations. Results show that, due to the fluctuations of allele frequencies between these generations their mutual information can only increase. Information $I(x,y)$ is also shown to constitute a new fixation index. Fixing occurs when a particular allele in a locus has reached a frequency equal 1. When it reaches a frequency of 0 it is lost. Interesting, we observed that the levels of $H_e$ of allozyme loci found in nature probably ensures the highest values of $I(x,y)$, in presence of at least a low level of genetic diversity. We also observed that, in absence of mutations, the mutual information over generations cannot be bigger that the entropy (genetic diversity) of the initial state ($\alpha$). Our results agree with Kimura's [15] theory in finding that neutral mutations are more effective in increasing the entropy of small populations.

## 2 Methodology

### 2.1 Discrete Event System

We suppose a genetic system composed by an initial state ($\alpha$) that represents the mother population and a final state ($\beta$), which is made up by the entropy of the offspring plus the mutual information between the offspring and mother population ($\alpha$). For simplicity, we will consider just one locus undergoing allele transitions $x_i \rightarrow y_j$. The occurrence of alleles $x_i$ in a locus $X$ at generation $t$ - 1 is determined by the probabilities $P(x_i)$ in the mothers ($\alpha$). The corresponding state ($\beta$) of alleles $y_j$ of a locus $Y$ at time $t$, occurs according the conditional probabilities $P(y_j/x_i)$ in the offspring, $j = 1,2, \ldots, n$; $i = 1, 2, \ldots, n$.

Let $\mathbf{X} = [p(x_1), \ldots, p(x_n)]$ denote a column vector made up by the probabilities $P(x_i)$ in the $\alpha$, and $\mathbf{C} = [P(y_j/x_i), j = 1,2, \ldots, n; i = 1, 2, \ldots, n]$ denote the *conditional event matrix*. The dispersal between mothers and samples (offspring) may be described by $P(y_j/x_i)$, that a gene sampled in given offspring $j$ had its parents in the mother population $i$ one unit time before. The vector of conditional probabilities, $P(y_j/x_i)$, must to be represented in a circular fashion [1, 19, 20]. Thus, $\mathbf{C} = circ\{C\}$ is a *circulant* matrix, defined by its first row $\mathbf{C_1}$, which is the vector $P(y_j/x_i)$, $j = 1,2, \ldots, n$. For brevity this is denoted as $p_j$, $j = 1,\ldots, n$. The next rows are cyclic permutations of the first one:

$$\mathbf{C} = \begin{bmatrix} p_1 & p_2 & \cdots & p_n \\ p_n & p_1 & \cdots & p_{n-1} \\ \vdots & \vdots & \ddots & \vdots \\ p_2 & p_3 & \cdots & p_1 \end{bmatrix} \quad (1)$$

by the law of total probability, $\mathbf{Y} = [P(y_1), \ldots, P(y_n)]$ defining the probabilities $P(y_j)$ of the $\beta$ state obeys

$$P(y_j) = \sum_{i=1}^{n} P(y_j|x_i)P(x_i) \tag{2}$$

the vector form is:

$$\mathbf{Y} = \mathbf{CX} \tag{3}$$

In this point is important to highlight that the entropy $H(y)$ of the $\beta$ state represents the total entropy of the system, which is made up by addition of measures of entropy of the offspring $H(y|x)$ plus the mutual information $I(x,y)$ between the mother population and the offspring [1].

The entropy of probabilities $p_i$, $i=1,\ldots n$ is

$$H(p) = -\sum_{i=1}^{n} p_i \log_b p_i. \tag{4}$$

The nature of base $b$ is discussed below. Likewise we replace $p_i$ by $P(x_i)$ to get the entropy $H(x)$ of the α state, with similar replacements by alternatively $P(y_j|x_i)$ and $P(y_i)$ to get the entropies $H(y|x)$ and $H(y)$, respectively. The mutual information $I(x,y)$ between states ($\alpha$) and ($\beta$) is

$$I(x,y) = H(y) - H(y|x). \tag{5}$$

More details about these entropies and the mutual information can be found at Shannon's [1] original work.

In a genetic system, the unit for measuring information is the level of ploidy. As usual, we choose base $b = 2$, giving $H$ and $I$ the convenient unit of 'bits.' Note that this assumes the presence diploids. In the case triploids the base $b = 3$ would be appropriate; etc. Most generally, the choice of $b$ connotes the size of the alphabet describing the system [1].

The entropies of a set of loci are *additive*, if there are $L$ loci present then Eq. (4) must be increased by factor $L$. This follows because, in the absence of gene linkage, the second law of Mendel postulates the independent segregation of loci, and in general, independent events obey additivity.

2.2 Shannon's Entropy as a Measure of Genetic Diversity

The expected proportion of heterozygotes per locus per individual ($H_e$) is a measure that summarizes the genetic variation in the level of allele. This parameter is often called *genetic diversity*. It obeys

$$H_e = 1 - \sum p_{ij}^2 \qquad (6)$$

where $p_{ij}$ is the frequency of the *i*-th allele of the *j*-th locus in a sample [21]. Both, $H_e$ and Shannon's entropy $H$ are natural measure of diversity, and functions of the space of possible outcomes. These reach their maximum values only in equiprobable cases, i.e. where $P(x_i) = 1/n$ for all $i=1,…,n$. Conversely they attain minimum values (in fact zero) when $P(x_i) = 1$ for some one $i$ and $P(x_{i'}) = 0$ for all $i' \neq i$ [1, 8, 21, 22].

The effective number of alleles ($n_e$) is another measure of genetic diversity, which has similar aspects to heterozygosity ($H_e$), and as well to $H$. As long, the magnitude of $n_e$ is also a function of the space of possible outcomes and depends on the degree to which the allele frequencies are even within populations or species. Rare alleles contribute little to the sum of the actual number of alleles $n_e$. It can also be taken as a measure of genetic information [21]. Its value is

$$n_e = \frac{1}{\sum p_i^2}. \qquad (7)$$

Taking the logarithm of $n_e$ to get an upper limit of $H$ of the genetic system

$$\log_b n_e = -\log_b \sum p_i^2. \qquad (8)$$

The upper limit of $n_e$ is reached when all the events are equiprobable. Then $n_e$ is found equal to the total number of alleles at one locus. Also, the upper limit of Shannon's entropy $H$ is equal to the logarithm of the number of events ($n$):

$$0 \leq H \leq \log_b n. \qquad (9)$$

Thus,

$$\log_b n_e = \log_b n. \qquad (10)$$

Substituting Eq. (10) in Eq. (9), we have the lower and upper limits the genetic entropy $H(J)$ to a given locus $J$:

$$0 \leq H(J) \leq \log_b n_e. \qquad (11)$$

2.2.1 A Genetic Entropy Measure

Suppose one locus and two alleles and $A$ and $a$, with allele frequencies $f(A) = p$, $f(a) = q$, and $p + q = 1$. The expected number of heterozygotes obeys $2pq$ [21]. Taking the logarithm of $H_e$ we find that this is equal to the *auto-information* proposed by Hartley [23]. Since $H$ is the *average* mutual information as defined by Shannon [1], it follows that

$$(H_e - H)^2 = \sigma^2, \qquad (12)$$

then directly

$$\sqrt{(H_e - H)^2} = \sigma. \qquad (13)$$

This result confirms $H$ as a genetic diversity measure, and this is equal to the standard deviation of allele frequencies plus the expected heterozygosity in one generation. Therefore, we defined a new concept of entropy that we called genetic entropy

$$G = \sigma + 2pq. \qquad (14)$$

The change in allele frequencies per generation is equal to binomial variance, $\sigma^2 = pq$ [14], then directly

$$G = \sqrt{pq} + 2pq. \qquad (15)$$

Observe in Table 1 that the measures of genetic diversity obtained as proposed by Shannon's entropy equation and the one obtained using the heterozygosity show approximate values (see last columns), for all allele frequencies. Observe that entropy and heterozygosity are genetic diversity measures that share a dependence upon the binomial variance. This would be important to say since it emphasizes that a purely genetic term, heterozygosity, relates to a Shannon information quantity 'the entropy'.

**Table 1**. Entropy measures by using the genetic parameters expected heterozygosity and standard deviation, and as proposed by Shannon's equation, and Kullback-Leibler distance between those entropy measures.

| $P$ | $H_e = 2pq$ | $\sqrt{pq}$ | $G = \sqrt{pq} + 2pq$ | $H = -(p\log_2 p + q\log_2 q)$ | $D(p||q)$ |
|---|---|---|---|---|---|
| 1 | 0 | 0 | 0 | 0 | 0 |
| 0.9999 | 0.000199 | 0.009999 | 0.010199 | 0.001473 | 0.004112 |
| 0.9996 | 0.000799 | 0.019996 | 0.020796 | 0.005092 | 0.010336 |
| 0.999 | 0.001998 | 0.031607 | 0.033605 | 0.011408 | 0.017780 |
| 0.9 | 0.18 | 0.3 | 0.48 | 0.468996 | 0.015692 |
| 0.8999 | 0.180159 | 0.300133 | 0.480293 | 0.469313 | 0.015659 |
| 0.8996 | 0.180639 | 0.300533 | 0.481172 | 0.470262 | 0.015559 |
| 0.899 | 0.181598 | 0.301329 | 0.482927 | 0.47126 | 0.016626 |
| 0.8 | 0.32 | 0.4 | 0.72 | 0.721928 | -0.002785 |
| 0.7999 | 0.320119 | 0.400075 | 0.720195 | 0.722128 | -0.002792 |
| 0.7996 | 0.320479 | 0.4003 | 0.720779 | 0.722727 | -0.002814 |
| 0.799 | 0.321198 | 0.400748 | 0.721946 | 0.723924 | -0.002856 |
| 0.7 | 0.42 | 0.458258 | 0.878258 | 0.881291 | -0.004383 |
| 0.69999 | 0.4200079 | 0.458262 | 0.87827 | 0.881303 | -0.004383 |
| 0.69996 | 0.4200319 | 0.458275 | 0.878307 | 0.88134 | -0.004382 |
| 0.699 | 0.420798 | 0.458693 | 0.879491 | 0.882510 | -0.004363 |
| 0.6 | 0.48 | 0.489898 | 0.969898 | 0.970951 | -0.001519 |
| 0.5999 | 0.4800399 | 0.489918 | 0.969958 | 0.971009 | -0.001516 |
| 0.5996 | 0.4801596 | 0.489979 | 0.970139 | 0.971184 | -0.001508 |
| 0.599 | 0.480398 | 0.490101 | 0.970499 | 0.971533 | -0.001491 |
| 0.5 | 0.5 | 0.5 | 1 | 1 | 0 |

$P$ = Frequency of allele $p$; $H_e$ = expected heterozygosity; $\sqrt{pq}$ = standard deviation of binomial variance; $G$ = genetic entropy; $H$ = Shannon entropy; $D(p||q)$ = Kullback-Leibler distance.

Kullback-Leibler distance, $D(p||q) = \sum p \log_2 \frac{p}{q}$, is a measure of the inefficiency given by assuming that the probability distribution is $q$, where $p$ is the true distribution. In this equation, we assume $q$ and $p$ indicate the measures of Shannon's entropy ($H$) and genetic entropy ($G$), respectively. As shown in the last column, those distances are almost zero. Entropy measures having the same actual and expected distributions are associated inefficiency zero. Therefore, we show that is also possible to calculate the genetic entropy measure to any population by using just genetic patterns as the variance and the standard deviation of allele frequencies, with no need for computing *log* functions. Note that the values of inefficiency larger than zero correspond to events that occur in the true distribution with higher probability than the expected distribution. Already events with inefficiency lower than zero are events that occur in the true distribution with a lower probability than the expected distribution.

2.3 The Measurements in Samples

We can represent the information-theoretical measures $H(Y)$, $H(Y/X)$, $I(X,Y)$ and $\bar{H}_e$ over many samples ($S$) from different populations, by taking the averages of those measures per generation

$$H(Y) = \frac{1}{S}\sum_{s=1}^{S} H_s(y), \tag{16}$$

$$H(Y|X) = \frac{1}{S}\sum_{s=1}^{S} H_s(y|x), \tag{17}$$

where $S$ is the total number of samples per generation ($s = 1, ...,S$). The mutual information is then

$$I(X,Y) = H(Y) - H(Y|X). \tag{18}$$

The average expected homozygosis per generation is

$$J_S = \frac{1}{S}\sum_{s=1}^{S} p_{ijs}^2, \tag{19}$$

here $p_{ijs}$ is the frequency of the $i$-th allele ($i = 1,..., n$) in the $j$-th locus ($j = 1,…, n$) of the $s$-th sample. Therefore, the average expected heterozygosity per generation is

$$\bar{H}_e = 1 - J_S. \tag{20}$$

2.4 Measurement of Real Diploid Data

In order to use these ideas to study genetic diversity is required knowledge of the allele frequencies from samples $s$ (offspring) and from the mother population ($\alpha$). The frequency of occurrence of $i$-th allele in the offspring at generation $t$ is conditional upon the frequency of this allele in the previous generation $t$-1 [24]. Thus, $P(y_j|x_i)$ is the number of copies of an allele in a generation $t$ divided by the total number of all alleles at the locus. As an example, if $n_A$ is a number of copies of the allele $A$ in a single sample $s$, its frequency is $p_A = n_A/2N_s$, where $2N_s$ is the total number of alleles in the diploid. In studies of genetic diversity, usually the allele frequencies of the total population (mother) are estimated taking the mean allele frequencies over the $S$ samples, $\bar{p}_i = \frac{1}{S}\sum_{s=1}^{S} p_{ijs}$ [14]. The average allele frequencies $\bar{p}_i$ are equal in the allele distribution of the ($\alpha$) state [24]. As described above, we applied our ideas to a medicinal plant (*Stryphnodendron adstringens*) from the Brazilian savanna; we used the estimates of allele frequencies from 627 individuals sampled in 16 populations [25]. We obtained the entropies and information measures of the natural and simulated genetic systems by using the software GENES, available free at http://www.genetica.dbg.ufv.br.

2.5 Mutual Information to Populations Undergoing Mutations

A new mutant allele that arises in a finite population will eventually be fixed or lost in it [15]. According to Kimura's model [16] a neutral mutation (not undergoing selection) in particular comes in only one copy, so its initial frequency is $p = 1/2N$. That is, in a population of fixed size, where the expectation of progeny is the same for all individuals in the population (effective number), the probability of a new silent mutation being fixed is inversely proportional to the population size ($2N$).

By using the entropy measure, we also found that neutral mutations are more effective for increasing the genetic diversity of small populations [16]. As an example, suppose a simplified situation where we have a mother population ($\alpha$) with two alleles *A* and *a*, showing the respective allele frequencies $p$ and $q$ in the generation $t$ -1. Suppose that the allele $q$ is undergoing recurrent mutation to a new allele called $r$. At the generation $t$ we have three offspring populations from $\alpha$ showing different sizes namely, $2N = 8$, $2N = 1000$ and $2N = 1.000.000$ (Table 2).

**Table 2**. Measures of entropies, mutual information and heterozygosity for offspring populations of different sizes undergoing neutral mutations.

|  | *t* -1 | *t* | | |
|---|---|---|---|---|
| **Allele frequencies** | $p = 0.5$<br>$q = 0.5$<br>$r = 0$ | $p = 0.5$<br>$q = 0.4$<br>$r = 0.1$ | $p = 0.5$<br>$q = 0.499$<br>$r = 0.001$ | $p = 0.5$<br>$q = 0.499999$<br>$r = 0.000001$ |
| $2N$ | $2N$ | 8 | 1000 | 1.000.000 |
| $H(x)$ | 1 | 1 | 1 | 1 |
| $H(y\|x)$ | - | 1.405 | 1.01 | 1.00002 |
| $I(x,y)$ | - | 0.1412 | 0.4904 | 0.49998 |
| $H(y)$ | - | 1.5462 | 1.5004 | 1.5000004 |
| $H_e$ | 0.5 | 0.58 | 0.501 | 0.500001 |

Note the comparison, in Table 2, the genetic diversity of the mother population and their offspring. These shows a small increase on diversity $H(y/x)$ to sample sizes from 1.000 to 1.000.000. The heterozygosity $H_e$ and mutual information $I(x,y)$ also are practically the same for the offspring showing these biggest sizes. On the other hand the smallest population ($2N = 8$) shows the highest increase for all the measures, compared to $\alpha$ and to the others offspring sizes.

Interestingly, the mean time for a rare mutant gene to become fixed in a finite population is proportional to its entropy [15]. Excluding the cases in which the allele is lost from it, a time fixation $\bar{t}(p)$ of an allele with frequency $p$ is [26]

$$\bar{t}(p) = -4N[p \log_2 p + (1-p) \log_2(1-p)]. \tag{21}$$

Thus,

$$\bar{t}(p) = -4N(H). \tag{22}$$

If a locus consists of a single allele, e.g., $A$, and a single mutation $a$ arise in a population, the fixation time to $a$ is

$$\bar{t}\left(\frac{1}{2N}\right) = 2 + \log_2 2N. \tag{23}$$

## 3 Simulations

Our aim is to show the effect of the fluctuations in the alleles transitions upon the entropy, mutual information and heterozygosity between a given generation $t$-1, represented by mother population, and the generation $t$, represented by the offspring. For simplicity we assume one locus and two alleles ($A$, $a$). These have respective frequencies $f(A) = p$ and $f(a) = q$, where $p + q = 1$. The mating is random, the generations distinct and offspring sizes $N_s$ constant over the generations. Within each sample, all individuals contribute equally to the pool of gametes for the next generation. We take into account samples of two different sizes, which we classify as:

*Small sample size*: we simulate the allele distributions of three $\alpha_i$ (mother populations) of size $N = 300$ with initial allele frequency $p_i$ equals 0.50, 0.67 or 0.95. For each such case we generate random offspring from $t = 1$ to $60$ generations. Each such generation $t$ consists of an allele sample taken at random from a large pool of zygotes from gametes produced in the previous generation $t - 1$, where the number of offspring (sample) is $S = 20$, and sample sizes $N_s = 20$.

*Big sample size:* we next simulate an ($\alpha$) state of size $N = 1000$, with initial allele frequency $p \approx 0.5$. We generate random samples of size $N_s = 300$ through $t = 21$ generations.

We use Eq. (3) to get the probabilities of the (β) state. We use the Eqs. (3) - (6) to get the entropy measures, mutual information and heterozygosity $H_e$, per sample per generation. Then Eqs. (16) - (20) are used to get these measures averaged per generation.

In a third simulation, we consider an ideal genetic system where mother population and offspring show no genetic differentiation, that is $P(x_i) = P(y_j|x_i)$, between the generations $t$-1 and $t$, respectively. We use 200 subintervals of length 0.005 for the $P$'s, starting at $P(x_i) = P(y_j|x_i) = 0$ and ending at $P(x_i) = P(y_j|x_i) = 1$.

## 4 Results and Discussion

### 4.1 Genetic Sampling Process

Studies of genetic diversity of natural populations especially focus upon estimating how different a current population, at generation $t_n$, is from an ancestral population ($\alpha_0$), at generation $t_0$. This is after intermediary generations $t = t_1, \ldots, t_{n-1}$. Accordingly in Fig. 1*a-c*, we show the fluctuations in the measures of entropy, mutual information and heterozygosity due to genetic differentiation through $n = 60$ offspring generations from an initial mother population ($\alpha_0$).

Given a fixed entropy $H(X)$ of ($\alpha_0$), the ensuing entropies $H(Y)$, $H(Y|X)$ and $\bar{H}_e$ tend to increase or decreased together, while $I(X,Y)$ has the opposite behavior (Fig. 1*a-d*). It is important to point out that we here show *average* entropic quantities over a set of samples. However, the same results hold for a unique sample. When the genetic system starts with equal allele frequencies, $p = q$, the entropies of ($\alpha$) and ($\beta$) states are equal, $H(X) = H(Y)$, and keep their maximum values over the generations $n$ (Fig. 1*a*). In general, the shapes of all the curves $H(Y|X)$, $\bar{H}_e$ and $I(X,Y)$ are sensitive to initial allele frequencies. In particular, the speeds of dispersion of the curves of $H(Y|X)$, $\bar{H}_e$ and $I(X,Y)$ toward their extremes are so sensitive. These are faster, e.g., when the system starts with $p_i = 0.50$ (Fig. 1*a*), and slower when instead $p_i = 0.95$ (Fig. 1*c*).

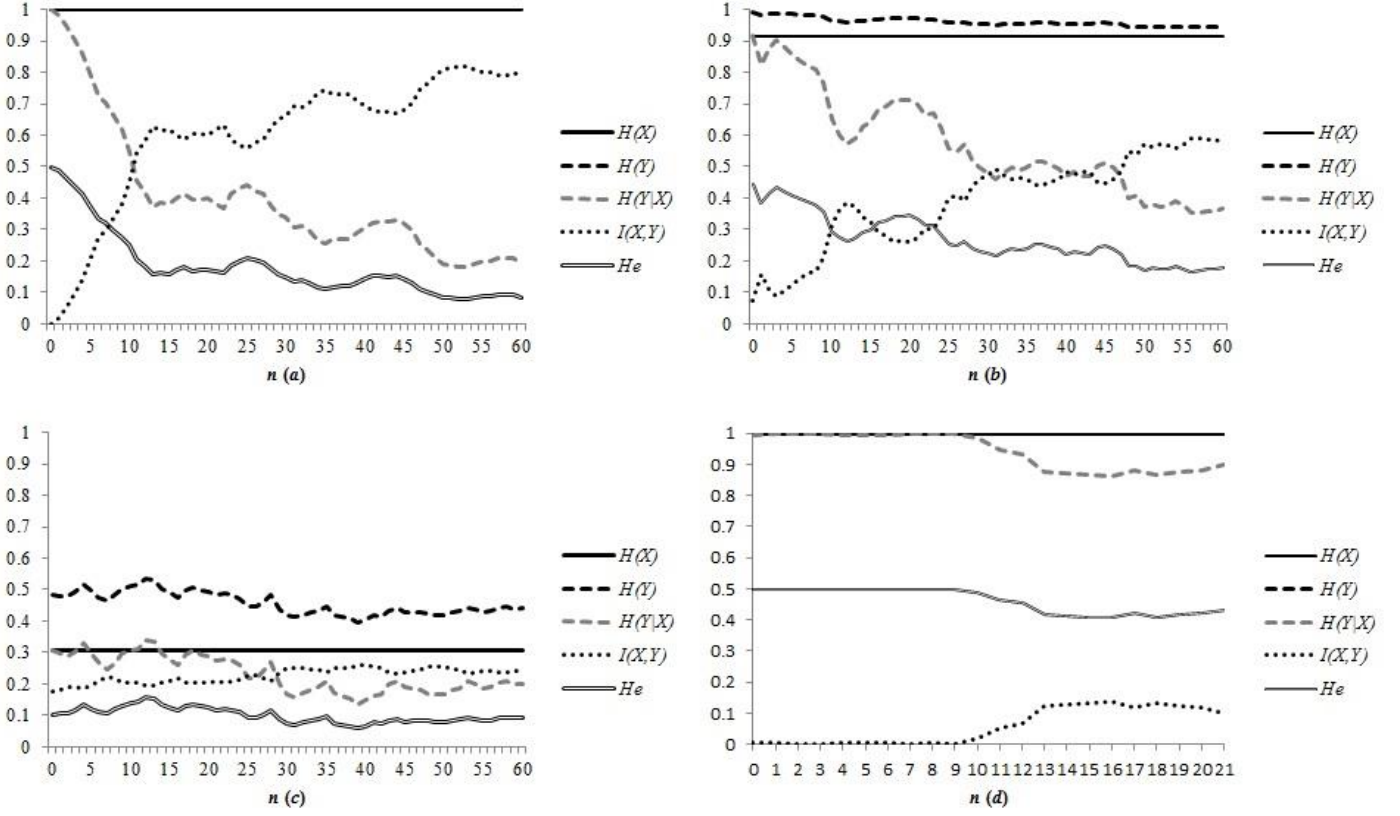

**Figure 1.** Entropies $H(X)$ of the α state, $H(Y)$ of the β state, $H(Y|X)$ of the samples (offspring); mutual information $I(X,Y)$ and expected heterozygosity ($\bar{H}_e$) of the genetic systems simulated over $n$ generations: *a - c*) small samples; *d*) big samples.

The chance fluctuation (noise) of allele frequencies due to small samples size is a well-studied process. The graph of the solution of Kimura [27] for Fisher equation [28] shows that the fixation time for samples from homogenous initial states ($p = q$), is relatively smaller. The duration of the dispersive phase depends on the initial allele frequencies distribution $P(x_i)$ and the sample size $N_s$. Through time, the samples become progressively differentiated until, eventually, the allele frequencies reach the limits of either 0 or 1 [24]. In those limits where the samples are showing fixed alleles [1]

$$H(X) = I(X,Y) = H(Y). \qquad (24)$$

The first equality shows that the 'intergenerational' information $I(X,Y)$ increases through time until it reaches its maximum value, the entropy $H(X)$ of the mother (initial state α). Note that this is not for merely the *sample*-averaged information but, rather, the direct Shannon information. In addition, the second equality shows that at the end of the process (final state β) the system has the *same entropy* $H(Y)$ as its value $H(X)$ at the beginning. This implies increase on order and hence decrease in genetic diversity (entropy) in absence of new mutations.

Wright [29] used a quantity $F_{IS}$ to designate the rate of approach toward fixation, in a finite population it is zero under random mating, or 1 under fixation. Here we observe that $I(x,y)$ is also a fixation index and it depends upon the probabilities $P(y_j|x_i)$. The information is zero for equal probabilities $P(y_j|x_i) = 1/n$ for all $i = 1,..., n$, and thus independent from the mothers ($\alpha$); or a maximum when $P(y_j|x_i) = 1$ for one index pair $i,j$ [1].

Bose [30] applied a similar approach to study the roles played by ecological processes in the maintenance of biodiversity. His results agrees with our allele results (*a*) and (*b*) of Fig. 1 showing that $I(X,Y)$ increases when diversity $H(Y/X)$ decreases.

4.2 Mutual Information between Undifferentiated Populations

We simulated an ideal situation where the mother population and the offspring show no genetic differentiation; $P(x_i) = P(y_j|x_i)$. That means absence of accumulation of differences in allele frequencies between populations due to evolutionary forces, such as selection or genetic drift. Fig. 2 shows the entropy of the mother population $H(x)$ and the total entropy of the system $H(y) = H(y_j|x_i) + I(x_i,y_j)$. We represent 200 subintervals of length 0.005 for the $P$'s starting at $P(x_i) = P(y_j|x_i) = 0.005$ and ending at $P(x_i) = P(y_j|x_i) = 1$. Therefore, there is no differences in allele frequencies between the mother and the offspring, but in each step (subinterval) they have their allele frequencies increased by the amount 0.005 until it reaches 1.

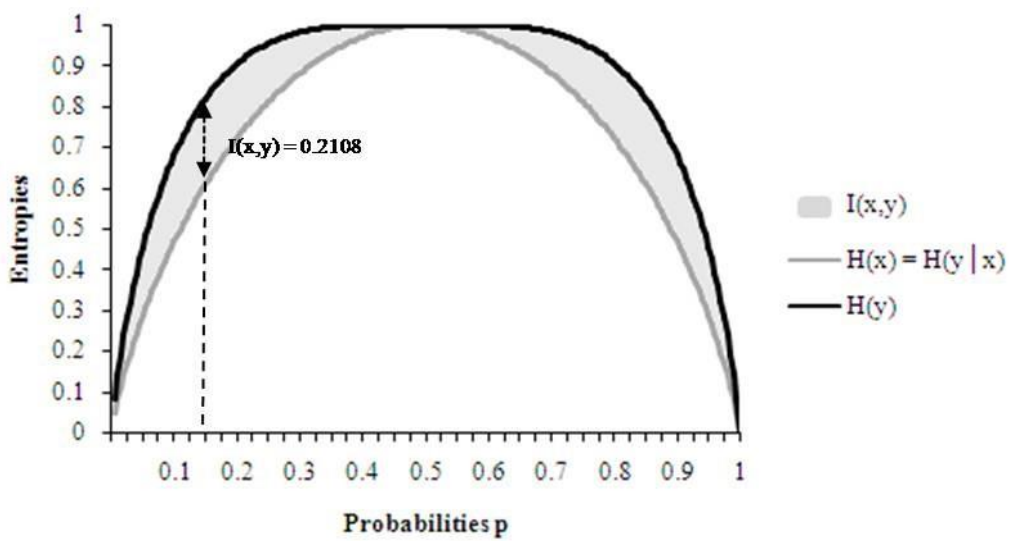

**Figure 2.** Measures of entropy of the mother population $H(x)$, offspring $H(y|x)$ and final state $H(y)$. And, mutual information $I(x,y)$ to populations showing no differences in allele frequencies.

In this simulation we also use on locus and two alleles. This allows us use the regular notation of one locus and two alleles $A$ and $a$. With allele frequencies $f(A) = p$, $f(a) = q$, and $p + q = 1$, observe that any increase in $p$ hence decrease $q$, and the reverse is true. The mother population and offspring starts with equal allele frequency, $p = 0.005$.

Recall that information $I(x,y)$ is defined to obey $I(x,y) = H(y) - H(y/x)$ [1]. In the current example, since mothers and offspring share equal allele frequencies, $H(x) = H(y|x)$. Therefore from the preceding $I(x,y) = H(y) - H(x)$. However, the total entropy $H(y)$ of the system can never be smaller than that of the mother population $H(x)$, from this, $H(y) \geq H(x)$. At first glance, this result seems somewhat unrealistic but, as we saw, it follows simply from basic information theory [1]. It is important to note that despite the inequality $H(y) \geq H(x)$ when the allele frequencies of the mother population are closer to the equivalence than the offspring, $H(x) > H(y/x)$. As well when the allele frequencies of the offspring are closer to the equivalence than the mother population, $H(x) < H(y/x)$.

Observe that as $H(x) = H(y|x)$ here, the difference between $H(y)$ and $H(y|x)$, that is the grey area between these line measure in the Fig. 2, is equal to $I(x,y)$. Note that the allele frequency $p$ grows from approximately zero (0.005) up 1. When $p = 0.5$ and $q = 0.5$, we found the maximum entropy values. When the mothers and the offspring shows very low levels of genetic diversity, $p \approx 0$ or $p = 1$, the system shows entropy or mutual information in fact zero ($p = 1$) or almost zero ($p \approx 0$). In the presence of any level of genetic diversity, $0 < p < 1$, we find the maximum values of $I(x,y)$ (gray area) when the allele frequencies are intermediaries between equivalence and the unit (or zero) and the equality ($p = q$). The vertical arrow in the Fig. 2 represents de difference between $H(y)$ and $H(y|x)$, and this is the highest value of $I(x,y) = 0.2108$ finding in presence of genetic diversity $H(y|x) \neq 0$, which represents allele frequencies $p = 0.14$ (dash line) and $p = 0.86$, by symmetry. Experiments in information theory to measure algorithmic complexity or algorithmic information contend (AIC) found that to the effective complexity to be sizable, the AIC must be neither too orderly nor too disorderly [31]. In addition, it was found that $I(X,Y)$ between successive generations is fairly stable for cases when biodiversity occurs. It was also found that $I(X,Y)$ shows an increasing trend for cases when biodiversity is diminished, meaning loss of species [30].

4.2 Mutual Information and Genetic Diversity of Natural Populations

The expected heterozygosity $H_e$ can be understood as a measure of diversity, homogeneity or heterogeneity. In this sense, low levels the heterozygosity imply highest mutual information levels. Thus, the measures $I(x,y)$ and $H_e$ might be used to describe levels of homogeneity of the allele distribution, when all outcomes are equally present in a sample of independent realizations of an experiment; and a measure of heterogeneity when all observations produce the same outcome. We pooled together results of many studies and those works shows that, the distribution of the allele frequencies to allozyme loci in natural populations is quite heterogeneous. Hamrick *et al.* [32] pooled data of 665 plant species into categories such as habits and life cycles and observed values of $H_e$ between 0.11 and 0.18. Loveless [33] compiled data from 97 species of tropical trees, classified them according to the mode of distribution, and observed values of $H_e$ between 0.12 and 0.20. Therefore, according to life cycles and distribution, species exhibit characteristic patterns of $H_e$. In studies using allozyme markers, Berg and Hamrick [21] reported typical values of $He$ between 0.04 to 0.25 for plants, from 0.0 to 0.17 for vertebrates, and from 0.0 to 0.28 for humans. The average $H_e$ found to *S. adstringens* (Table 3) was 0.1681 [25].

Hartl and Clark [34] found that, in general, the observed heterozygosis values for allozyme loci of most organisms range 0.04 to 0.14. This heterogeneity in the distribution (1- 0.04 = 0.96 to 1–0.14 = 0.86) has pointed as an indication that the null hypothesis may be wrong. This would imply in the selection of allozyme loci, which would be a premature conclusion since electrophoresis routine procedures do not distinguish all alleles. On the other hand, the allozyme polymorphism may be overestimated because allozyme typically investigated are those found in relatively high concentration in tissues or body fluids. Despite of controversy about the excess of heterogeneity, or not, found to allozyme loci in nature, these low levels of $H_e$ implies alleles frequencies around 0.85, and probably ensures the highest values of $I(x,y)$.

It is common knowledge that the decrease in genetic diversity is dangerous to most species. In some cases, this might have catastrophic consequences by causing an "excess" of order as represented by Eq. (24). In Fig. 1 analysis, we presumed the absence of mutations. Evidently, these are essential for the evolution of increased diversity in species (See also Sec. 2.5). However, even with a highly dominant allele in neutral locus as observed in many species (above mentioned), this "apparent" heterogeneity probably "hide" high levels of genetic diversity. For example, in the Fig. 2 the highest level of $I(x,y) = 0.2108$ represents only 26.47% of the total entropy at this point, $H(y) = 0.7964$, while the genetic diversity of the offspring $H(y/x)$ is equal to 73.53% of this value.

Table 3 shows the average measures of entropies, $I(X,Y)$ and $\overline{H}_e$ to 6 isozyme loci from 16 populations of *S. adstringens*. The isozyme systems were disposed in decreasing order according to the proportion of decreasing of fixed alleles, what means increasing of diversity. The measures $H(X)$, $H(Y)$, $H(Y/X)$, $\overline{H}_e$ and $I(X,Y)$ increase in value as the proportion of allele fixation decreases (except for systems of EC code numbers *2.6.1.1* and *1.1.1.25,* as seen in the Table 3 very tiny difference of 4.9x10$^{-3}$). This means that increasing the amount of the entropy in the genetic systems also increases the amount of information. $I(X,Y)$. The inequality $H(Y) \geq H(X)$ was observed in all isozyme systems and for all populations (individual data not shown), and as expected $I(x,y)$ obeyed the limits

$$0 \leq I(x,y) \leq H(x). \qquad (25)$$

**Table 3.** Average measures of entropies, mutual information and heterozygosity for 16 populations of *S. adstringens* to six isozyme loci. $H(X)$ = entropy of the initial state ($\alpha$); $H(Y)$ = entropy of the final state ($\beta$); $H(Y|X)$ = entropy of the samples; $I(X,Y)$ = mutual information; $\bar{H}_e$ = heterozygosity expected.

| $^{EC}$ Isozymes | % Fixation | Allele number | $H(X)$ | $H(Y)$ | $H(Y|X)$ | $I(X,Y)$ | $\bar{H}_e$ |
|---|---|---|---|---|---|---|---|
| *1.1.1.1* | 70 | 2 | 0.0986 | 0.1532 | 0.0659 | 0.0873 | 0.0234 |
| *2.6.1.1* | 59 | 3 | 0.1455 | 0.2322 | 0.1046 | 0.1276 | 0.0347 |
| *1.1.1.25* | 41 | 3 | 0.1507 | 0.2472 | 0.1245 | 0.1227 | 0.0394 |
| *5.3.1.9* | 29 | 2 | 0.4427 | 0.5421 | 0.2189 | 0.3232 | 0.0797 |
| *1.1.1.37* | 0 | 3 | 0.6205 | 0.9244 | 0.5589 | 0.3654 | 0.2181 |
| *1.1.1.42* | 0 | 5 | 1.8323 | 2.1537 | 1.5747 | 0.5791 | 0.6135 |
| Average | | | 0.5484 | 0.7088 | 0.4412 | 0.2675 | 0.1681 |

$^{EC}$ = Enzyme Commission code number.

## 5 Conclusions

Shannon's entropic measures $H(y/x)$ and $I(x,y)$ are suitable tools for predicting the state of genetic diversity in natural populations. We found that $I(x,y)$ between successive generations is quite stable for cases when the system is neither too orderly ($p \approx 1$) nor too disorderly ($p \approx 0.5$) (Fig. 1*c*). Thus, $I(x,y)$ shows an increasing trend for cases when genetic diversity is diminished by loss of alleles, being a measure of fixation of alleles.

However, $I(x,y)$ is also a measure of the information transmitted from one generation to the next. It is maximized as value $I(x,y) = H(x)$ under allele fixation. Thus, the transmitted information cannot be greater than the genetic diversity of the initial mother population, in the absence of new mutations.

**Acknowledgments:** The first author thanks the University of Arizona, and the 'Science without Borders' scholarship program for grants funded by the Brazilian Federal Government, 'Coordenação de Aperfeiçoamento de Pessoal de Nível Superior (CAPES)' and 'Conselho Nacional de Desenvolvimento Científico e Tecnológico (CNPq)'.